\documentclass[12pt]{article}
\usepackage{epsfig}
\newcommand{\eq}{\begin{equation}}
\newcommand{\eqx}{\end{equation}}
\newcommand{\eqn}{\begin{eqnarray}}
\newcommand{\eqnx}{\end{eqnarray}}
\newcommand{\dt}{\Delta}

\newcommand{\dd}{\partial}
\newcommand{\lab}{\label}
\begin{document}
%
%\begin{titlepage}
%\pagestyle{empty}
%\vspace*{2cm}
\begin{center}
{\large\bf QCD predictions for polarised deep inelastic scattering 
accompanied by a forward jet in the low $x$ region of possible HERA 
measurements} 
\vspace{1.1cm}\\
{\sc J.~Kwieci\'nski,}\footnote{e-mail: jkwiecin@solaris.ifj.edu.pl}
{\sc B. Ziaja}\footnote{e-mail: beataz@solaris.ifj.edu.pl}\\
\vspace{0.3cm}
{\it Department of Theoretical Physics,\\
H.~Niewodnicza\'nski Institute of Nuclear Physics,
Cracow, Poland}
\end{center}
%\vspace{0.6cm}
%
\begin{abstract}
We estimate the cross-section and asymmetry relevant for 
the production of forward jets in polarised deep
inelastic scattering in the region of small values of $x$ 
which can be probed at possible polarised HERA measurements. 
The kinematical cuts implemented in our analysis are the same 
as those which were used in the unpolarised deep inelastic scattering at HERA. 
The calculations are based on  the double $ln^2(1/x)$ resummation  
which controls the polarised deep inelastic scattering for small 
values of the Bjorken parameter $x$.    
We show that this resummation  substantially enhances  
the corresponding cross-section and asymmetry.  The predicted value 
of the asymmetry is found to vary between $-0.01$ and $-0.04$ within 
the small $x$ region which can possibly be probed at polarised 
HERA measurements.  
\end{abstract}
%\end{titlepage}
%\vspace{1.1cm}
%%%%%%%%%%%%%%%%%%%%%%%%%%%%%%%%%%%%%%%%%%%%%%%%%%%%%%%%%%%%%%%%%%
The  polarised deep inelastic scattering at small $x$ is dominated 
by the effects of the double $ln^2(1/x)$ resummation \cite{BARTNS,BARTS,
BZIAJA2}.  Here, as usual, $x=Q^2/(2pq)$, where $Q^2=-q^2$ with $q$ and $p$ 
denoting the four momentum transfer between leptons and four momentum of the proton 
respectively.     
It has been pointed out in \cite{BZJKJET} that, apart from the structure 
function measurement, a process which probes the effects of 
double $ln^2(1/x)$ resummation might be the deep inelastic scattering 
accompanied by a forward energetic jet. The idea of studying the deep inelastic $ep$ scattering with an identified
forward jet as a probe of the low $x$ behaviour of QCD was proposed
by Mueller \cite{MUELLER}. Since then it has been successfully 
applied to the 
case of unpolarised deep inelastic scattering (DIS) 
at HERA \cite{FJET1,FJET2,FJET3}.   
The Mueller's proposal was
to study the unpolarised deep inelastic events $(x,Q^2)$ containing
an identified forward jet $(x_J,k^2_{J})$, where 
 the longitudinal momentum fraction $x_J$ of the proton carried by
a forward jet and jet transverse momentum squared $k^2_{J}$ 
were assumed to fulfill the conditions~:
\eqn
x_J & \gg & x,\lab{assumx}\\
k^2_{J} & \sim & Q^2\lab{assumk}.
\eqnx

The first condition $ x_J  \gg  x$ guarantees that one probes the QCD dynamics
in the low $x/x_J$ region (the quantities which are measured depend
on the ratio $x/x_J$). The second condition $k^2_{J} \sim Q^2$
supresses the effects 
of the standard leading order (LO) DGLAP evolution (from $k^2_{J}$ to
$Q^2$), and so one can probe the low $x$ QCD effects which go beyond this evolution.  
In the unpolarised DIS the forward jet measurement is therefore regarded as
a useful tool for probing the BFKL pomeron \cite{BFKL}.  Experimental results 
on forward jet production  in deep inelastic scattering at HERA are presented 
in refs.\ \cite{JETH1,JETZEUS}.    

The BFKL pomeron decouples from polarised deep inelastic  scattering. At low
$x$ the latter is dominated by the novel effects of the double $ln^2(1/x)$ 
resummation.  The dominant contribution to this resummation is generated by ladder 
diagrams with quark (antiquark) and gluon exchange(s) along the ladder.  
The double logarithmic $ln^2(1/x)$ terms come from the configuration of strongly 
ordered $x_n$ and $k_n^2/x_n$ \cite{QED},  where $k_n^2$ and  $x_n$ denote the transverse
momenta squared and longitudinal momentum fractions of the proton exchanged 
along the parton ladder respectively.   
The effects of the double 
$ln^2(1/x)$ resummation should therefore be still visible in the kinematical 
configuration $k_J^2 \sim Q^2$. The effects of the conventional (LO) QCD evolution 
from the scale $k_J^2$ to $Q^2$ are on the other hand expected 
to be suppressed in this configuration. 

In ref.\ \cite{BZJKJET} we have discussed in detail the differential structure 
function $x_J{\partial^2 g_1\over \partial x_J \partial
k_j^2}$ describing (forward) jet production.  It is linked to the corresponding differential cross-section 
in the standard way:
\eq
{\dd^4 \sigma \over \dd x \dd Q^2 \dd x_J \dd k_J^2}=
\frac{8\alpha_e^2\pi}{Q^4}\,\,\,
{\dd^2 g_1 \over \dd x_J \dd k_J^2}
\,\, y (2-y),
\lab{cross}
\eqx
where $\alpha_e$ denotes the electromagnetic coupling constant, and $y$ describes
the energy fraction of incoming electron carried by the interacting virtual 
photon, i.e. $y=qp/(p_ep)$ where $p_e$ denotes the for momentum 
of the incoming electron.  
In equation (\ref{cross}) we have omitted terms proportional to $\gamma^2 = 
4M^2x^2/Q^2$, where $M$ denotes nucleon mass,  which are negligible 
at small $x$. 
The cross section in formula (\ref{cross}) corresponds, as usual, to the 
difference between the cross-sections for antiparallel and parallel spin 
orientations \cite{SMC}.  Similarly as for the unpolarised case, restrictions (\ref{assumx}),
(\ref{assumk}) applied to the polarised DIS forward jet events
allow one to neglect the effects of DGLAP evolution
and to concentrate on the low $x/x_J$ behaviour of
$x_J{\dd^2 g_1 \over \dd x_J \dd k_J^2}$. 

In this note we wish to study the cross-section itself, together with the 
asymmetry, taking into account the kinematical cuts which have recently been 
used in the forward jet measurements in unpolarised deep inelastic 
scattering at HERA \cite{JETH1,JETZEUS}.  To be precise we shall give predictions for the cross-section 
$d\sigma /dx$~: 
\eq
{d\sigma \over dx} = \int dk_J^2 dx_J dQ^2  
{\dd^4 \sigma \over \dd x \dd Q^2 \dd x_J \dd k_J^2}
\lab{incx}
\eqx
with the integration region restricted by the kinematical cuts which  will be 
specified below.  The asymmetry $A(x)$ is conventionally defined as: 
\eq
A(x) = {d\sigma \over dx}\,/\,\left(2\,\,  {d\sigma_u \over dx}\right),
\lab{asym}
\eqx
where  ${ d\sigma_u \over dx}$ is the cross-section for forward jet production 
in unpolarised deep inelastic scattering.

The formula for the differential structure function can be written in
the following form (see Fig.\ 1)~:
$$
x_J{\dd^2 g_1\over \dd x_J\dd k_J^2}
= 
$$
\begin{equation}
{<e^2> \over 2} \bar \alpha_s(k_J^2)\sum_{im }\Delta p_i(x_J,k_J^2)\Delta P_{mi}(0)
[ \Phi_{m}^{S}({x\over x_J},k_J^2,Q^2) +
 \Phi_{m}^{NS}({x\over x_J},k_J^2,Q^2)],  
\label{dg1}
\end{equation}
%
%FIGURE 1
\begin{figure}[t]
    \centerline{
     \epsfig{figure=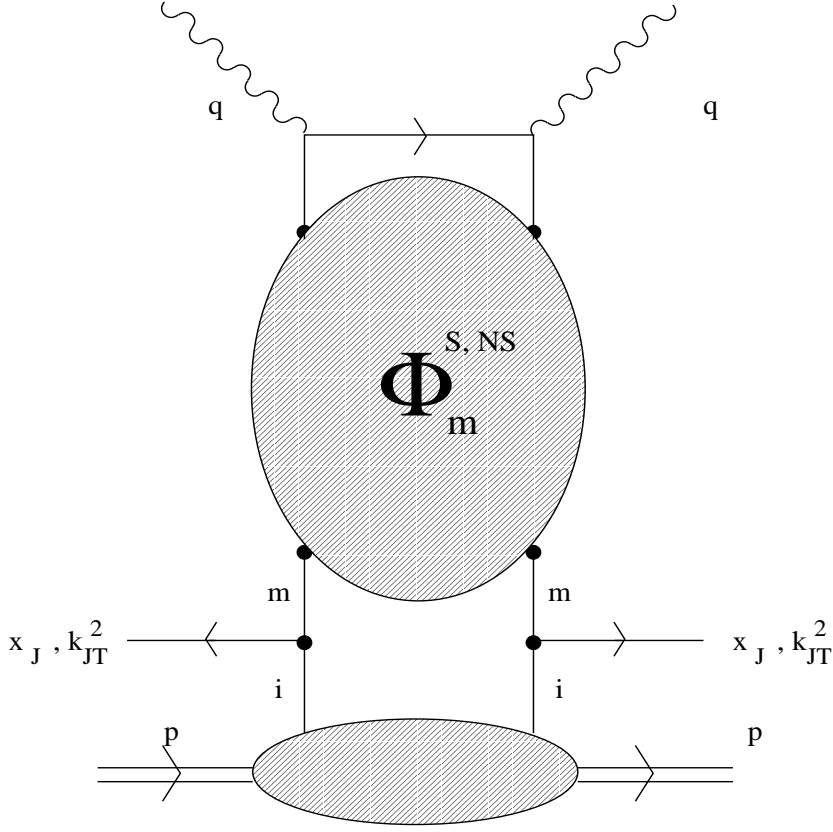,height=11cm,width=11cm}
               }
\caption{Diagrammatic representation of the formula (\ref{dg1}) 
for the differential structure function 
$x_J{\partial^2 g_1\over \partial x_J \partial k_J^2}$ describing the 
forward jet production in polarised deep inelastic scattering.}
\label{fig.1}
\end{figure}
\noindent
where $\Delta p_i(x_J,k_J^2)$ are the (integrated) spin dependent parton distributions
in the proton.  The quantities $\Delta P_{im}(0) \equiv \Delta P_{im}(z=0)$,
where $\Delta P_{im}(z)$ (for quarks $\Delta P_{im}(z)=\delta_{im}\dt P_{qq}$)
denote the LO splitting functions describing evolution
of spin dependent  parton densities. The indices $i$ and  $m$ numerate quarks, antiquarks
and gluons and the functions $\Phi_{m}^{S,NS}({x\over x_J},k_J^2,Q^2)$ 
correspond to the singlet and non-singlet combinations of the 
spin dependent structure functions of the parton $m$.  The average quark charge 
squared $<e^2>$ is defined as: 
\begin{equation}
<e^2> = {1\over N_f} \sum_{i=1}^{N_f} e_i^2,
\label{ave2}
\end{equation}
where $N_f$ denotes the number of quark flavours which we set equal to 3.  
Finally, $\bar \alpha_s = \alpha_s/2\pi$. Equation (\ref{dg1}) was derived,  
assuming  strong ordering of transverse momenta 
($k_i^2 << k_J^2 \sim k_m^2$) and of the longitudinal momentum fractions 
($x_m << x_i \sim x_J$) at the jet production vertex (see Fig.\ 1).  

\noindent
The functions $\Phi_{m}^{S,NS}({x\over x_J},k_J^2,Q^2)$ are related to the 
singlet and the non-singlet combinations of 
unintegrated  quark and antiquark distributions
$f_{m}^{S,NS}({\bar x\over x^{\prime}}, k_J^2, k_f^2)$ in the parton $m$,
where $ k_f^2$ denotes the transverse momentum squared of the quark (antiquark) 
which is probed by the virtual photon:
{\small
\eqn
\Phi_{m}^{S,NS}({x\over x_J},k_J^2,Q^2)=
\int_{x_{down}}^{x_{up}} {dx^{\prime}\over x^{\prime}}
\int_{k^2_{down}}^{k^2_{up}}
{dk_f^2\over k_f^2}
f_m^{S,NS}({\bar x\over x^{\prime}},k_J^2, k_f^2),
\label{sumphiml}
\eqnx
where $\bar x =x\left(1 + {k_f^2\over Q^2} \right)$.
The integration limits $x_{up}$, $x_{down}$, $k^2_{up}$ and
$k^2_{down}$~:
\eqn
x_{up}=x_J,& & x_{down}=x(1+{k_J^2\over Q^2}),\nonumber\\
k^2_{up}=Q^2(\frac{x'}{x}-1),& & 
k^2_{down}=k_J^2/(\frac{x'}{x}-\frac{k_J^2}{Q^2})
\lab{limits}
\eqnx
take into account restrictions implied by the phase-space limitation 
$k_f^2 < Q^2 (x_J/x-1)$ and   
by the ordering of longitudinal momentum fractions $x_n$ and of the  
ratios $k_n^2/x_n$ along the ladder diagrams defining the functions 
$f_m^{S,NS}(\xi,k_J^2, k_f^2)$  \cite{BZJKJET}.  
To be precise, the functions $f_m^{S,NS}(\xi,k_J^2, k_f^2)$   
are the solutions of the integral equations which correspond 
to ladder diagrams in the double logarithmic $ln^2(1/\xi)$ approximation.  
The functions $f_m^{NS}(\xi,k_J^2, k_f^2)$, with $m$ corresponding to 
$u, \bar u, d,\bar d$ are generated by ladder diagrams with quark (antiquark) 
exchange along the chain.  The singlet combinations $f_m^{S}(\xi,k_J^2, k_f^2)$
($m=u, \bar u, d,\bar d,s,\bar s$) mix with the functions
$f_m^{g}(\xi,k_J^2, k_f^2)$  describing the 
(spin dependent) gluon distribution in  parton $m$.  Both functions are 
generated by coupled integral equations corresponding to quark (antiquark) 
and gluon exchanges along the ladder. The integral equations for
the functions $f_m^{S,NS,g}(\xi,k_J^2,k_f^2)$ resumming the
double $ln^2(1/\xi)$ terms are given in ref.\ \cite{BZJKJET}.  
The solutions of the 
integral equations for  $f_m^{S,NS,g}(\xi,k_J^2, k_f^2)$ depend upon 
the scale $\mu^2$ which  defines the coupling 
$\bar \alpha_s(\mu^2)$  controlling the elementary vertices along 
the partonic ladder. In ref.\ \cite{BZJKJET} we have considered two limiting cases 
which allow the analytic solution, i.e. $\mu^2=(k_J^2 + Q^2)/2$ and $\mu^2=
k_f^2/\xi$. For the former choice of the scale the coupling does not change
along the ladder while the latter case corresponds to the
 coupling which varies along the chain.  The analytic expressions for the functions 
$f_m^{S,NS}(\xi,k_J^2,k_f^2)$  for both cases are given 
in ref.\ \cite{BZJKJET} .
  
\noindent
In order to calculate the cross-section $d \sigma/dx$ we have used the following
kinematical cuts:~ 
\begin{table}[hbpt]
\noindent
\centerline{Tab.\ 1}
\begin{center}
\begin{tabular}{||c||}
\hline \hline
$E_e^{\prime} > 11 GeV$\\
\hline
$y > 0.1$\\
\hline
$160^0 < \theta_e < 173 ^0$\\
\hline
$x_J > 0.035$\\
\hline
$k_J > 3.5 GeV $\\
\hline
$7^0 < \theta_J < 20^0 $\\
\hline
${Q^2\over 2}< k_J^2 <  2\,Q^2$\\ 
\hline \hline
\end{tabular}
\end{center}
\end{table}
%%%%%%%%%%%%%%%%%%%%%%%%%%%%%%%%%%%%%%%%%%%%%%%%%%%%%%%%%%%%%%%%%%%%%%%%

\noindent
We also set $E_e = 30 GeV$ and $E_p = 820 GeV$. The variables 
$E_e, E_p, E_e^{\prime}, 
\theta_e$ and $\theta_J$ denote the energy of the incoming electron, 
the energy of the incident proton, the energy of the scattered electron, the 
scattering angle of the electron and the angle of the measured jet respectively.  
The variables $x_J$ and $k_J^2$ denote as before the longitudinal momentum 
fraction $x_J$ of the proton carried by
a forward jet and jet transverse momentum squared $k^2_{J}$.   
The angles $\theta_e$ and $\theta_J$ are measured with respect to the direction of the incident proton.      
The  cuts summarised in Table 1 are the same as those which were used by the H1 collaboration 
in the analysis of forward jet production in unpolarised deep inelastic 
scattering \cite{JETH1} in the small $x$ region which can possibly also 
be probed at the polarised HERA measurements \cite{ALBERT}.  

In Figures 2a - 2c we show our predictions for the spin dependent cross-section 
$d\sigma /dx$ plotted as the function of $x$.  We present results corresponding to two choices of the scale 
$\mu^2$,  i.e. $\mu^2 = (k_J^2 + Q^2)/2$  (Fig.\ 2a) and 
$\mu^2 = k_J^2/\xi$ (Fig.\ 2b)  (the scale $\mu^2$ is the argument 
of the QCD coupling controlling the kernels of the 
integral equations for the functions
$f_m^{S,NS,g}(\xi,k_J^2,k_f^2)$). In  Fig.\ 2c we show our result
which was obtained by neglecting the double $ln^2(1/\xi)$ 
resummation effects.   This approximation  corresponds to neglecting the 
parton radiation between the forward jet and the virtual photon.
It may be seen from the results shown in Figures 2a-2c 
that the impact  of the double $ln^2(1/\xi)$ 
resummation effects on the cross-section is very significant.  
These effects give the cross-section 
which rises much  more steeply with decreasing $x$, down to 
$x \sim 5\cdot 10^{-4}$,  than the
cross-section which 
corresponds to the case where the double $ln^2(1/\xi)$ effects
were neglected.     
At their  maximum the two cross-sections differ by a factor
equal to about 20 or 40, depending upon the choice of the scale $\mu^2$.         
The decrease of the cross-section with decreasing $x$ in the
region $x <  5\cdot 10^{-4}$ is a kinematical effect 
which follows from the cuts (see Tab.\ 1).~ 
%
%FIGURE 2
\begin{figure}[t]
\centerline{\epsfig{figure=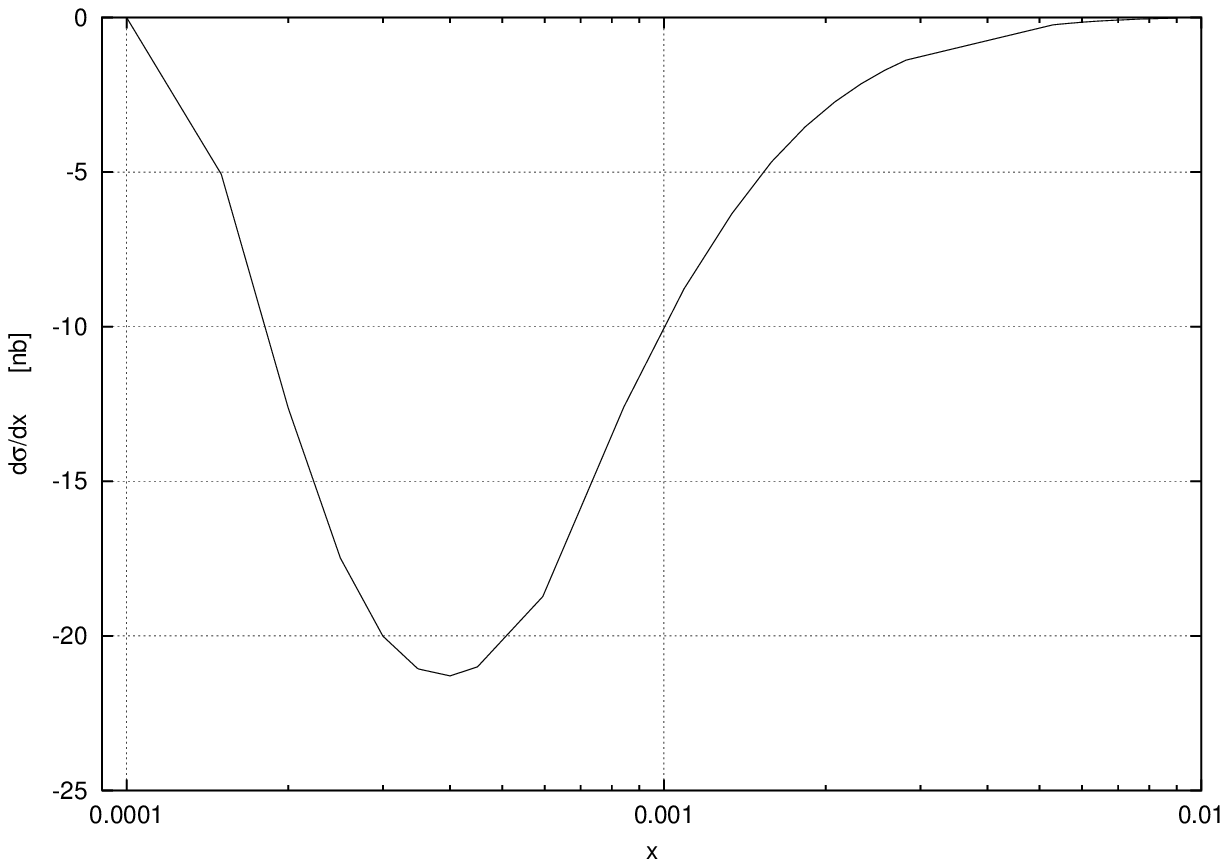,height=6cm,width=9cm}}
\centerline{\epsfig{figure=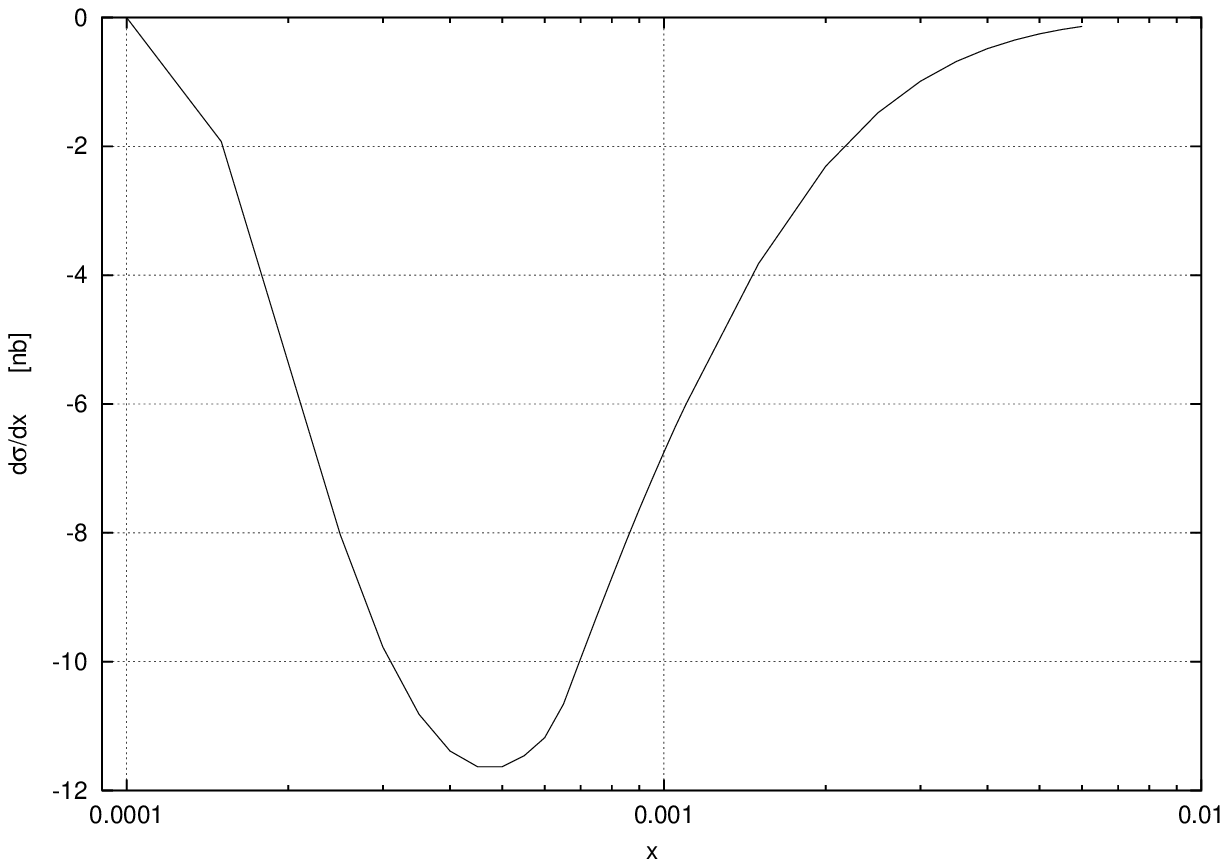,height=6cm,width=9cm}}
\centerline{\epsfig{figure=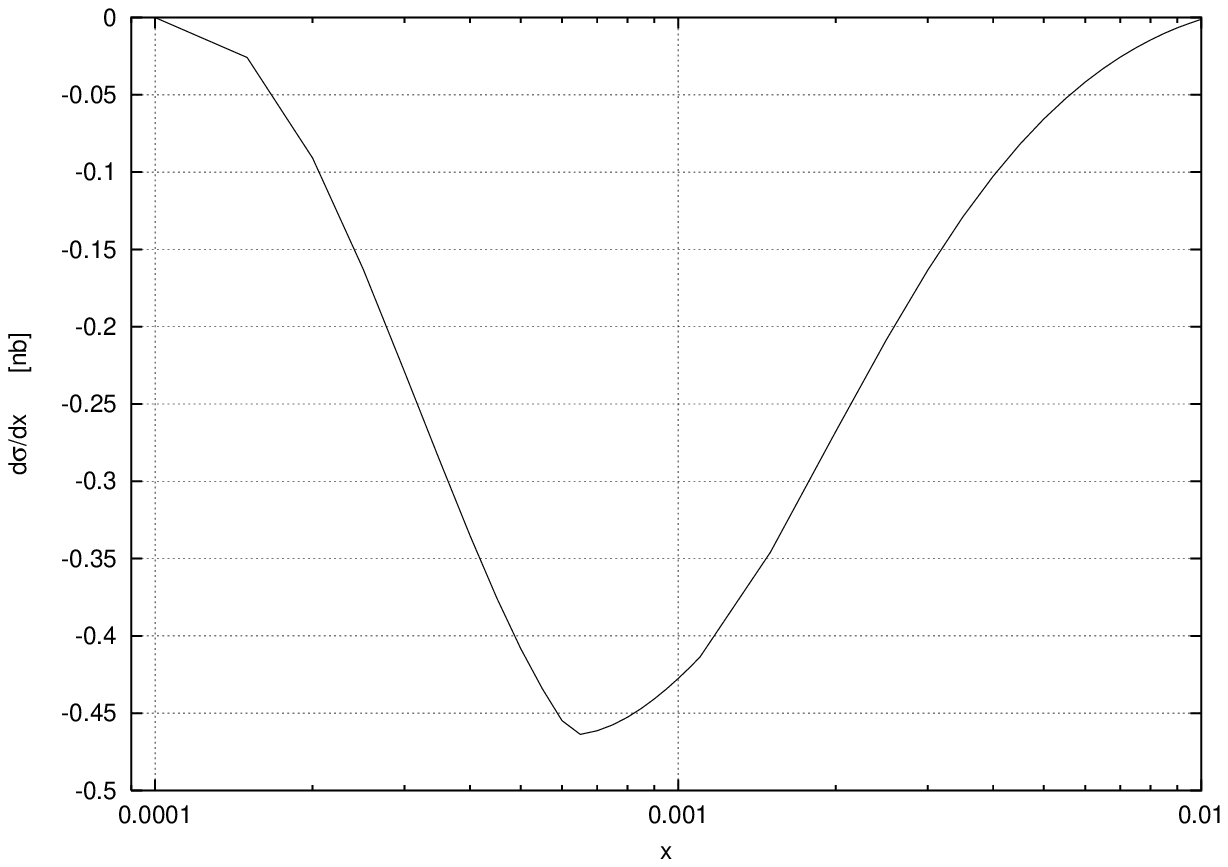,height=6cm,width=9cm}}
\caption{
The cross-section ${d\sigma \over dx}$ for forward jet production
in polarised deep inelastic scattering  plotted as the function of 
$x$.  Figures 2a and 2b show results which include the double
logarithmic $ln^2(1/\xi)$ effects and correspond to two 
 choices of the scale $\mu^2$ which appears as the argument of the
QCD coupling  controlling the kernels of the corresponding
integral equations  for the functions $f_m^{S,NS,g}(\xi,k_J^2,k_f^2)$.   
Fig.\ 2a corresponds to $\mu^2=(k_J^2 + Q^2)/2 $ and Fig.\ 2b to 
$\mu^2=k_f^2/\xi$ respectively.  
Fig.\ 2c shows the cross-section  ${d\sigma \over dx}$ in the approximation 
when the double logarithmic resummation is neglected.} 
\label{fig.2}
\end{figure}

In Fig.\ 3 we show  predictions for the asymmetry $A(x)$ defined by equation 
(\ref{asym}).  The asymmetry $A(x)$ was obtained by combining our results 
for $d\sigma /dx$ with the results of the calculation 
of the cross-section $ d\sigma_u /dx$ describing the jet production in 
unpolarised deep inelastic scattering with the same kinematical cuts as those 
summarised in Tab.\ 1 \cite{FJET3}.  
The presented asymmetry $A(x)$ weakly decreases 
with decreasing $x$.   It is found to be approximately equal to 
 $-0.01$ for $x = 0.004$ and to decrease down to  
$-0.038$ for $\mu^2 = (k_J^2 + Q^2)/2$ or to $-0.022$ 
for $\mu^2 = k_f^2/\xi$ at $x = 0.0005$.~
%
%FIGURE 3
\begin{figure}[t]
\centerline{\epsfig{figure=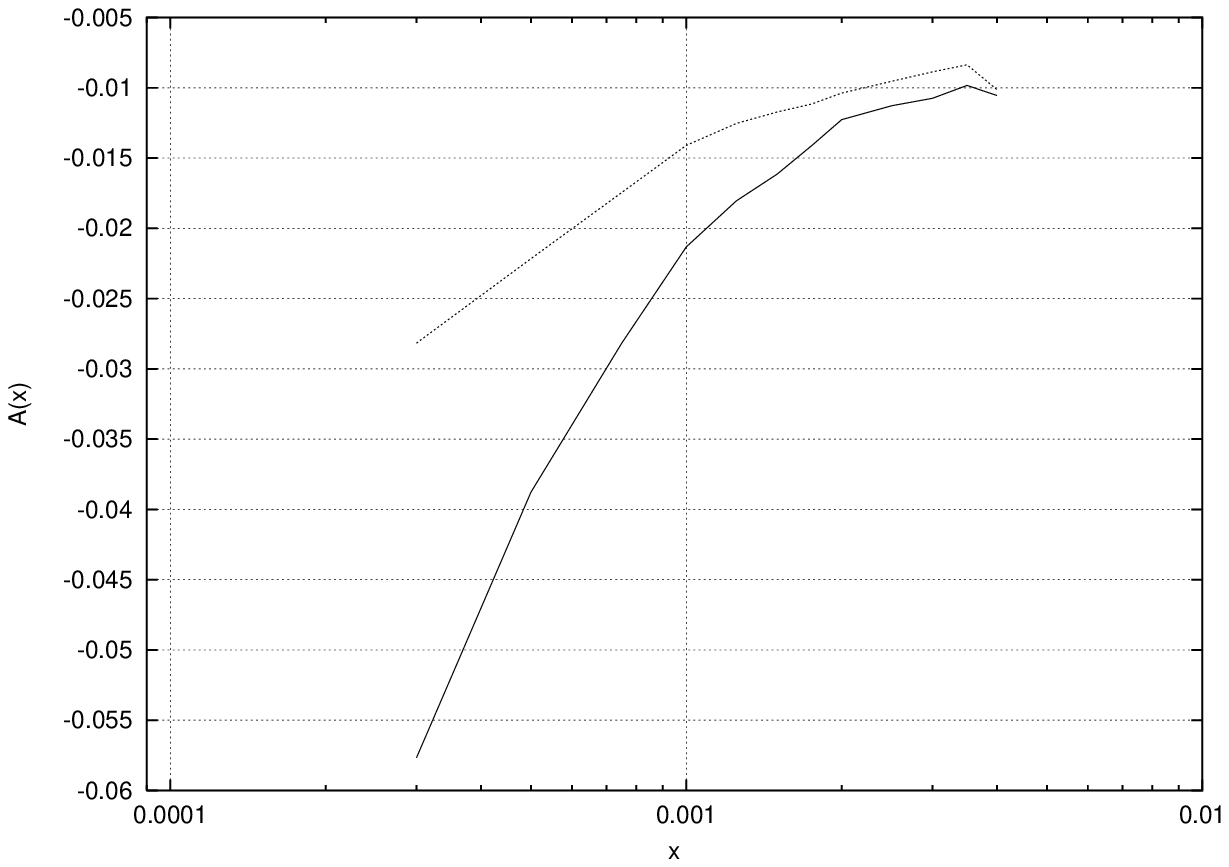,height=6cm,width=9cm}}
\caption{
The asymmetry parameter $A(x)$ defined by eq.\ (\ref{asym}).  
The two curves correspond to two choices of the  scale $\mu^2$ 
which appears as the argument of the QCD coupling  controlling 
the kernels of the corresponding integral equations for the 
functions $f_m^{S,NS,g}(\xi,k_J^2,k_f^2)$.  
The lower and upper curves correspond to  
$\mu^2=(k_J^2 + Q^2)/2 $ and to $\mu^2=k_f^2/\xi$ respectively.}
\label{fig.3}
\end{figure}

To sum up, we have estimated in this note the impact of the
double $ln^2(1/x)$ effects on the corresponding cross-sections 
and asymmetries describing the forward jet production in
polarised deep inelastic lepton scattering in the small $x$
regime 
which can be probed at the possible polarised HERA
measurements.  We have found that double logarithmic effects
substantially enhance the absolute magnitude of the
corresponding cross-section and of the asymmetry.   
These quantities would be negligibly small if the double
logarithmic 
effects were neglected.  The asymmetry parameter $A(x)$ was
found to vary between $-0.01$ and $-0.04$ at the small $x$ range
relevant for the possible polarised HERA measurements.

\section*{Acknowledgments}

%%%%%%%%%%%%%%%%%%%%%%%%%%%%%%%%%%%%%%%%%%%%%%%%%%%%%%%%%%%%%%%%%%%%%%%%%%
We thank John Outhwaite for providing us with the results of the calculation 
of $d\sigma_u /dx$. 
This research has been supported in part by the Polish Committee for Scientific
Research grant 2 P03B 89 13   and by the
EU Fourth Framework Programme 'Training and Mobility of Researchers', Network
'Quantum Chromodynamics and the Deep Structure of Elementary Particles',
contract FMRX--CT98--0194.

\end{document}